# Enhanced second harmonic generation from resonant GaAs gratings

D. de Ceglia[1,*], G. D'Aguanno[1], N. Mattiucci[1], M. A. Vincenti[1], M. Scalora[2]

[1]AEgis Technologies Inc., 401 Jan Davis Dr, Huntsville – AL, 35806, USA

[2]Charles M. Bowden Research Center, AMSRD-AMR-WS-ST, Redstone Arsenal, Huntsville, AL 35898, USA

*Corresponding author: ddeceglia@nanogenesisgroup.com

## Abstract

We study second harmonic generation in nonlinear, GaAs gratings. We find large enhancement of conversion efficiency when the pump field excites the guided mode resonances of the grating. Under these circumstances the spectrum near the pump wavelength displays sharp resonances characterized by dramatic enhancements of local fields and favorable conditions for second harmonic generation, even in regimes of strong linear absorption at the harmonic wavelength. In particular, in a GaAs grating pumped at 1064nm, we predict second harmonic conversion efficiencies approximately five orders of magnitude larger than conversion rates achievable in either bulk or etalon structures of the same material.

The conversion efficiency (CE) for second harmonic generation (SHG) depends on factors like phase and group velocity matching, and may be improved in several ways. One approach is to increase the coherence length of the interacting fields through phase matching and quasi-phase matching techniques, so that the nonlinear (NL) process is not limited by the interaction length [1-2]. One may also design high-quality resonant cavities for the harmonic fields to boost the local field amplitudes [3]. The CE can be significantly limited by linear absorption at the pump and second harmonic (SH) wavelengths. However, this need not necessarily be the case, provided one exploits the inhomogeneous solution of the wave equation, which corresponds to the phase-locked (PL) harmonic component.

The solution of the SH NL wave equation may be expressed as a superposition of homogeneous and inhomogeneous solutions, also known as free and bound waves [4]. The bound SH component is trapped and fully phase locked to the pump. While the homogeneous solution propagates according to the expected material dispersion, the PL component displays the same propagation properties (absorption and phase) as the fundamental wavelength [5]. Thus, the PL component survives even if the material at the SH wavelength is characterized by huge absorption, in either bulk or cavity environments [6], provided the pump is tuned in a region of relative transparency. We presently exploit another mechanism, based on Wood's anomalies of a diffraction grating made by perforating a slab of NL material (e.g. GaAs), to enhance the field localization of a pump signal at 1064nm and boost PL-SH CE at 532mn, where GaAs is mostly opaque.

Gratings are characterized by two types of Wood's anomalies [7-8]: (i) the Wood-Rayleigh anomalies, associated with sharp variations of diffraction efficiencies at the onset or disappearance of diffracted orders [9]; and (ii) resonance type anomalies, generated by diffracted orders that are phase-matched to resonant modes of the grating [10]. Guided mode resonances (GMRs) belong to the latter type of anomalies. The diffraction grating obtained by modulating the index of a waveguide along its propagation direction displays abrupt changes of diffraction efficiency whenever the propagation modes of the waveguide are excited. These anomalies appear regardless of the nature of the excited modes, which may be TE or TM-polarized, and can involve surface waves, like surface plasmon polaritons in metallic gratings, or core-guided modes. A notable aspect of this phenomenon is that spectral and angular bandwidths of the resulting resonances are usually very narrow, a few nanometers or less, making these gratings highly selective filters [11]. An example of this kind of anomaly is the narrow-bandwidth transmission resonance of one-dimensional (1D) metallic gratings under TM illumination, excited by selecting periodicities close to the SP wavelength [12].

According to references [13] and [14], SHG can be enhanced by exciting leaky eigenmodes in two-dimensional photonic crystal membranes, where both pump and SH fields are resonant. Similar phenomena have been observed in a resonant waveguide grating covered with a NL polymer [15]. The zeroth order, diffracted SH signal is enhanced by one order of magnitude as the pump excites the fundamental leaky grating mode. Enhanced SHG has also been demonstrated in all-dielectric gratings [16-17], where the quadratic NL process arises in the form of a weak surface effect induced by symmetry breaking at the interfaces.

In this Letter we study SHG at 532nm, i.e. the typical configuration where a Nd:Yag laser is used as the pump, in a 1D, free-standing, semiconductor grating with subwavelength slits. We exploit the grating periodicity to excite TE-polarized waveguide modes that form Fano-like resonances [18] in the diffraction efficiency spectrum near the pump field wavelength, both in reflection and transmission. We stress that the SH falls in the visible range, where linear absorption of GaAs yields an attenuation length of only 125nm, so the homogeneous SH component is absorbed after propagating that distance into or on the surface of the material. Therefore, any attempt to increase the CE based on phase-matching techniques would fail, since the necessary component is absorbed, leaving intact the PL component. Our study of the CE will involve two basic structures: (i) a GaAs bulk (a slab of GaAs embedded in a linear medium with matching refractive index) where the interacting fields propagate forward; (ii) a GaAs etalon surrounded by air that supports Fabry-Perot transmission resonances. We assume a TE-polarized pump field at 1064nm incident at a 10° angle to exploit the bulk quadratic susceptibility of GaAs, $\chi^{(2)}$=10pm/V, and we consider a TE-polarized SH field generated at 532nm. This typical experimental setup is described in reference [5].

We begin with a linear analysis of the grating. In Fig.1 we show the grating and the input pump at 10°.

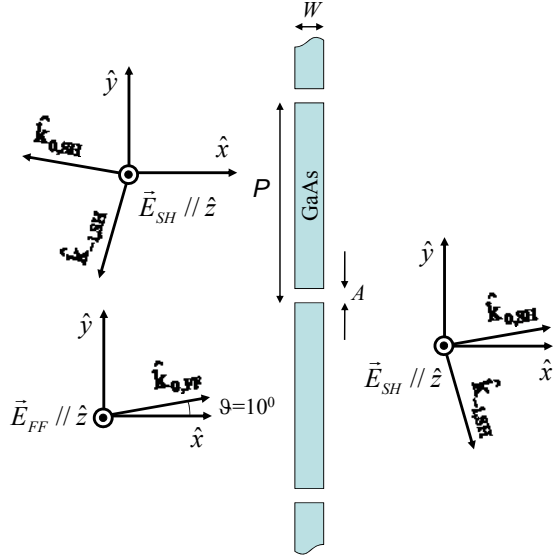

Fig. 1. Sketch of the TE pump signal incident at θ=10⁰ on the grating. The subscripts FF and SH refer to the pump and SH fields, respectively. For a pitch $P$=500nm (a case later analyzed in Fig. 4), the grating is subwavelength for the pump field (1064nm), while the diffracted SHG is distributed on the zeroth order at $\theta_{0,SH}$=10⁰ and the first order at $\theta_{-1,SH}$=−63⁰.

The power of the radiated fields (pump and SH) is measured in the forward and backward directions by integrating the x-component of the Poynting vector along the reflection and transmission sections, and by normalizing those values with respect to the input power per unit cell $P_{inc}$. In Fig.2 the transmission coefficient $T$ at 10° for the pump wavelength ($\lambda_{FF}$=1064nm) was mapped as a function of the pitch $P$ and the grating thickness $W$, fixing the slit size at $A$=32nm. In this map two resonant phenomena are clearly recognizable: (i) the wide resonances, whose positions are not altered by the periodicity $P$, are Fabry-Perot resonances due to multiple reflections at the input and output interfaces of the structure; (ii) the sharp resonances, much more dependent on $P$, are GMRs triggered by modes propagating along the y-direction, the grating periodicity axis. In this configuration, the air filling factor $A/P$ is much smaller than one so that the slits may be considered a tiny perturbation on the smooth slab waveguide and the GMRs are well approximated by the mode of the planar structure. The phase matching condition for the excitation of GMRs is $k_{GMR} \cong k_{WM} = |k_0 \sin(\vartheta_{inc}) + 2\pi m/P|$, $m = \pm(1,2,3,...)$ where $k_0$ is the wavevector of the incident plane wave, $\vartheta_{inc}$ is the angle of incidence, $k_{GMR}$ are the GMRs' wave-vectors, $k_{WM} = k_0 n_{WM}$ are the guided modes' wave-vectors, and $n_{WM}$ are the effective refractive indices of the guided modes supported by the slab waveguide. The electric field distribution for two GMRs is reported in Figs.2b-c; the similarity of these resonant states to the shape of a $TE_0$ mode and a $TE_1$ mode is apparent.

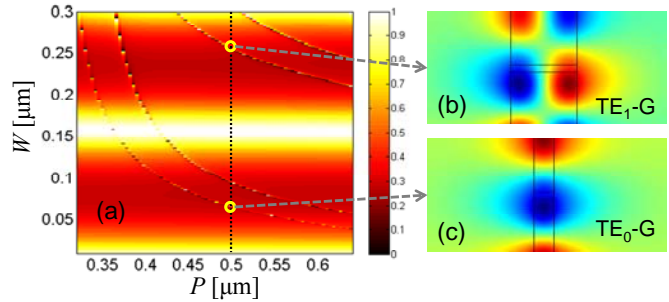

Fig. 2. (a) Transmission of the GaAs grating, at $\lambda_{FF}$=1064nm for a 10° incidence angle, as a function of the periodicity $P$ and thickness $W$. (b) Electric field localization at the GMR excited with a grating thickness $W$~250nm (see the circle in (a)). (c) Same as in (b) for a grating thickness $W$~66nm. $G=2\pi/P$ is the grating momentum.

In Fig. 3 we report the dispersion of the unperturbed modes as a function of the slab thickness $W$, along with the grating periodicity required to match the effective refractive indices of these modes for an incident parallel momentum $k_0 \sin(\vartheta_{inc})$. The comparison of these curves with the narrow GMRs traces reported in Fig.2 shows that the unperturbed modes are a good qualitative guide to predict the position and dispersion of GMRs.

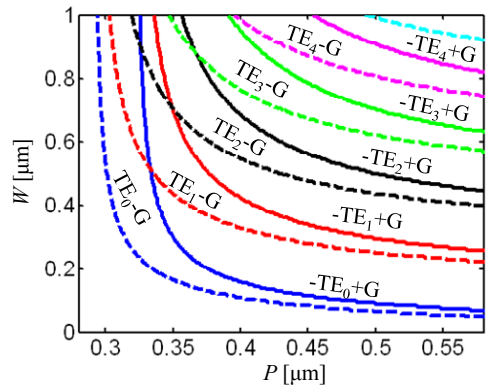

Fig. 3. Geometric dispersion of the guided modes in the absence of the slits array. The effective refractive indexes are analytically evaluated as the eigenmodes of a slab waveguide. The grating momentum is calculated at the pump wavelength 1064nm and for an incidence angle of $10^0$.

We now calculate the SH CE from gratings with periodicity $P$=500nm, aperture size $A$=32nm ($<<\lambda_{FF}$) and variable thickness $W$. The linear transmission of the gratings may be inferred by tracing the map in Fig.2 along the axis $P$=500nm. In Fig. 4 we compare the normalized forward SHG efficiency $\eta = P_{SH} / P_{FF}^2$ of this set of gratings to the efficiencies of a bulk structure, i.e. a single layer of GaAs impedance-matched to the adjacent medium, and an GaAs etalon layer embedded in air having the same thickness $W$ of NL material. In the bulk structure, the free SH component is fully absorbed in less than a coherence length $Lc=\lambda_{FF}/[4(n_{FF}-n_{SH})]$~390nm, where $n_{FF,SH}$ are the refractive indices at the pump and SH wavelengths, respectively. The CE is clamped at a constant value, a condition imposed by the presence of the phase locked SH signal. In the etalon the forward SH intensity is strongly correlated to the Fabry-Perot resonances for the pump field. Even in this structure, for cavities longer than few coherence lengths, the only significant SHG contribution comes from the PL component. The forward SHG efficiency of the grating overlaps well the etalon efficiency except when the pump signal couples to GMRs for specific grating thicknesses that yield huge conversion spikes. Backward and forward efficiencies are almost the same.

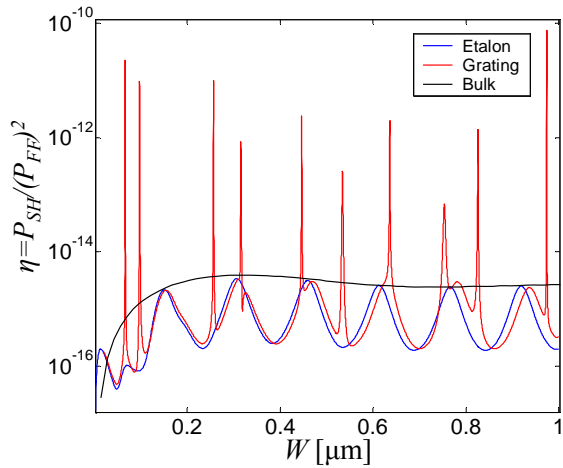

Fig. 4. Normalized SH CE of the 1D grating, the bulk and the etalon structures as a function of thickness.

The nature of the enhancement coincides with the strong field localization available for the pump field at the GMRs (see Figs. 2b-c). At the same time material absorption at the SH wavelength plays no role in the NL interaction, and similar conversion rates are possible for either thin (see the peak at $W$~66nm) or thick structures (see the peaks at $W$>0.5µm). In the same vein we observe that GaAs may be replaced with any other quadratic NL material with similar results. The choice of different filling factors or slit sizes will merely shift the spectral positions of GMRs without altering conversion efficiencies. In our calculations we used relatively small pump intensities (a few kW/cm$^2$) to avoid triggering $\chi^{(3)}$ processes, which in the presence of such narrow-band resonant states (on the order of one nanometer) are expected to induce switching and multi-stability phenomena at low-pump-intensity thresholds.

Finally, we calculated the distribution of the scattered SHG on the multiple diffraction orders available at the SH wavelength. The diffraction angles for the generated field follow the classical diffraction equation at the SH wavelength. As an example, for a grating with *P*=500nm and *W*~66nm, 80% of forward SHG efficiency is diffracted at the zeroth order angle ($\theta_{0,SH}=10^0$) and 20% at the angle $\theta_{-1,SH}=-63^0$ of the order *m=−1*.

In conclusion, we have demonstrated drastically enhanced SHG at 532nm in a 1D GaAs subwavelength grating. The coupling of the pump field to GMRs introduces sharp, Fano-like resonances where strong field localization and enhancement take place, leading to the prediction of conversion efficiencies that are four to five orders of magnitude larger than bulk or etalon structures. Moreover, thanks to the PL SH component the efficiency is not influenced by linear absorption at the harmonic wavelength, and is unrelated to grating thickness or to the order of the guided mode excited. A comment regarding the effect of the substrate is in order. For the sake of simplicity we have studied a free-standing grating. However, it is likely that the grating will be deposited on glass or similar substrate. The introduction of a substrate causes only red-shifts of the spectral positions of the GMRs due to the red-shift of the corresponding waveguide modes associated with the air/GaAs/substrate structure.